\documentclass[twocolumn,showpacs,preprintnumbers,amsmath,amssymb,prb]{revtex4}%

\usepackage{graphicx}
\usepackage{dcolumn}
\usepackage{bm}

\begin{document}

\preprint{}

\title{Kinetic Energy Density Study of Some 
\\
Representative Semilocal Kinetic Energy Functionals}

\author{David Garc\'{\i}a-Aldea}
\email{dgaldea@fisfun.uned.es}

\author{J. E. Alvarellos}
\email{jealvar@fisfun.uned.es}

\affiliation{Departamento de F\'{\i}sica Fundamental. UNED. }
\affiliation{Apartado 60.141. E-28080 Madrid (Spain). }
\date{\today}

\begin{abstract}
There is a number of explicit kinetic energy density functionals 
for non-interacting electron systems that are obtained in terms 
of the electron density and its derivatives. 
These semilocal functionals have been widely used
in the literature. In this work we present a comparative study 
of the \emph{kinetic energy density} of these semilocal 
functionals, stressing the importance of the local behavior 
to assess the quality of the functionals. 
We propose a \emph{quality factor} that measures the local 
differences between the usual orbital-based kinetic energy 
density distributions and the approximated ones, 
allowing to ensure if the good results obtained for the total
kinetic energies with these semilocal functionals are due 
to their correct local performance or to error cancellations. 
We have also included contributions coming from the laplacian 
of the electron density to work with an infinite set of 
kinetic energy densities. 
For all the functionals but one we have found that their 
success in the evaluation of the total kinetic energy are 
due to global error cancellations, whereas the local 
behavior of their kinetic energy density becomes worse 
than that corresponding to the Thomas-Fermi functional.

\end{abstract}

\pacs{31.15.Ew, 31.15.Bs, 31.15.-p, 71.10.CA, 71.15.Mb}
\maketitle

%Valid %PACS numbers may be entered using the \verb+\pacs{#1}+ command.


%------------------------------------------------------------------------
%body of paper here - Use proper section commands
%References should be done using the \cite, \ref, and \label commands

\section{\label{sec:intro}Introduction}

Density Functional Theory (DFT) has nowadays a privileged position within the
current methods for calculating the electron structure. The theorems of
Hohenberg and Kohn \cite{1964HK} show that the ground state of a system of
nuclei and electrons can be fully described in terms of its electron density.
In fact, the total energy $E[n]$ of the electrons can be written as a
functional of the density and the variational minimization of $E[n]$ yields
the ground state electron density and its total energy.

The usual procedure within the DFT solves the Kohn-Sham\cite{1965KS} (KS)
equations for $N$ orbitals, where $N$ is the number of electrons of the
system. This KS method allows the partition of the total energy functional in
different pieces with distinct physical meanings
\begin{equation}
E[n]=T_{S}[n]+V[n]+J[n]+E_{xc}[n],\label{eq:totE}%
\end{equation}
where $E[n]$ is the total energy, $V[n]$ is the energy of the electron density
allocated in the electric field generated by the nuclei, $J[n]$ is the
classical repulsion of the electron density (also called Hartree energy),
$T_{S}[n]$ is the kinetic energy of a noninteracting system that yields the
same electron density of the interacting one and $E_{xc}[n]$ is a density
functional that takes into account the exchange and the correlation (XC)
energies. The major part of the kinetic energy, $T_{S}[n]$, is then expressed
exactly by means of the KS one-electron orbitals and the small part 
($T[n]-T_{S}[n]$) is included in the XC functional, $E_{xc}[n]$, which is the
only part of the energy to be approximated. The exact ground-state 
electron density
$n\left(  \mathbf{r}\right)  $ is the sum of the densities of the KS orbitals,
the kinetic energy functional is evaluated exactly through the KS
orbitals,\cite{1965KS} and the minimization of the total energy functional
(\ref{eq:totE}) becomes the resolution of a set of coupled
Schr\"{o}dinger-like equations for the KS orbitals.

The computational cost of solving the KS equations can be avoided by using
$T_{S}$ as a functional depending explicitly on the electron density, instead
of constructing the KS orbitals and evaluating the kinetic energy in terms of
them. 
In that case, the computational cost of this \emph{orbital-free}
procedure would scale with the number of electrons (a linear scaling 
method) and can offer much faster computational times than the KS 
calculations, allowing to deal with systems that involve several hundred 
thousands of atoms or more.
The ground state of these systems is then calculated through an Euler-Lagrange
variational minimization. Moreover, the use of an approximate orbital-free
kinetic energy functional, instead of the KS method, makes easier the
evaluation of the forces and reduces the computational cost of complex
calculations in first-principle molecular dynamics \cite{1985CP}. But one of
the issues that must be solved about orbital-free kinetic energy functionals
is the stability of their numerical solutions, because some functionals have
been proved to be linearly stable but nonlinearly unstable (see, e. g., Ref.
\onlinecite{2005BC}) and the solution obtained can be meaningless.

As a consequence, the construction of functionals depending explicitly on the
density $n\left(  \mathbf{r}\right)  $ (without any reference to the wave
functions) has an undoubted formal interest and an important practical side.
Being the analytic dependence of $T_{S}$ on the electron density a rather
academic problem for a long time (see reviews in Refs.
\onlinecite{b1989ParrYang} and \onlinecite{r2000WangCarter}), the increasingly
rapid development of computational chemistry has turn it an interesting topic,
from the proposals of new kinetic energy
functionals\cite{1978AG,%
1985CAT,1996GAC1,1996GAC2,1998GAC1,1998GAC2,2000GAC,%
1992WT,1996FM,%
1999WGC,2000WC,2003CarlingCarter,2005Zhou}
to the study of the kinetic energy density\cite{1996YLW,2003SLBB} and the
application to simple systems.\cite{2001IEMS,2001CCH,2001KH}

In order to construct accurate explicit $T_{S}\left[  n\right]  $ functionals
we need to get not only good total energies but also the correct density
profiles for the ground state of the total energy functional
(Eq.~\ref{eq:totE}) after a fully variational minimization of the energy. For
that reason, although common tests to determine the quality of a kinetic
energy functional only calculate the energies and density profiles using
\emph{good} densities (i.e., those obtained with accurate methods such as the
Hartree-Fock or KS ones) we are interested in studying the quality of the
kinetic functionals beyond this limitation. In particular, we think the
properties of the kinetic energy density deserve a study by itself, paving the
way for understanding when and why functionals are able or not to describe the
characteristic quantum properties, like the presence of structure in the
density profiles.

This paper presents an extensive study of the kinetic energy densities of a
number of semilocal kinetic functionals thanks to the proposal of a quality 
factor that helps to determine how far from a valid kinetic energy density 
are the approximated ones.

\section{\label{sec:KED}The Kinetic Energy Density}

The kinetic energy density (KED) can be defined as any function $t_{S}%
(\mathbf{r})$ that integrates to the exact total kinetic energy,
\begin{equation}
T_{S}[n] = \int d\mathbf{r~}t_{S}(\mathbf{r}).
\end{equation}
It is clear that such definition does not determine the kinetic energy density
uniquely: any function -- with the appropriate scaling properties -- that
integrates to zero in the whole space can be added to any KED to yield another
KED, and an infinite number of new valid KEDs can be defined by multiplying
this function by any coefficient. The non uniqueness of the kinetic energy
density has been studied in the literature.\cite{1996YLW,2003SLBB}

Within the KS method, an orbital-based KED can be obtained in terms of the KS
orbitals. Two main definitions are commonly used in the literature. The first
definition has the advantage of being positive everywhere, because it is
calculated through the squared gradient of the orbitals of an 
$N$-electron system (atomic unit will be used in this paper):
\begin{equation}
t_{S}^{I}(\mathbf{r})=\frac{1}{2}\sum_{i=1}^{N}\left\vert \nabla\phi
_{i}(\mathbf{r})\right\vert ^{2},\label{eq:def1}%
\end{equation}
where $\phi_{i}(\mathbf{r})$ is the $i$th KS orbital. 
A second definition can
be obtained using the kinetic energy operator in the way it appears in the KS
equations, yielding a non positive definite function:
\begin{equation}
t_{S}^{II}(\mathbf{r})=-\frac{1}{2}\sum_{i=1}^{N}\phi_{i}^{\ast}%
(\mathbf{r})\nabla^{2}\phi_{i}(\mathbf{r}).\label{eq:def2}%
\end{equation}
Functions $t_{S}^{I}$ and $t_{S}^{II}$ are related through the laplacian of
the electron density, 
\begin{equation}
t_{0}(\mathbf{r})=\nabla^{2}n(\mathbf{r}),
\end{equation}
as
\begin{equation}
t_{S}^{I}(\mathbf{r})-t_{S}^{II}(\mathbf{r})=%
\frac{1}{4}t_{0}(\mathbf{r}).
\end{equation}
They have different local properties, but both of them are valid 
definitions for KED because they integrate to the same total 
kinetic energy since the
integral of $t_{0}(\mathbf{r})$ over the whole space is zero. 
In finite systems the electron density decays exponentially and 
the divergence theorem can be used to transform the integral over 
the space into a surface integral,
$\int_{V}d\mathbf{r\nabla}^{2}n(\mathbf{r})=%
\int_{S}d\mathbf{S\nabla}n(\mathbf{r})$.
The gradient of the electron density decays faster than the growth of the
surface (which is proportional to $r^{2}$) and the integral is zero. 
For extended systems, but periodic, the last integral is extended over 
the surface of the unit cell of the periodic system; 
the gradients in opposites sides of
the cell cancel one each other, and the surface integral 
has also a zero value.

We must remark that, besides the interest of the study of the intrinsic
properties of the KED, some new XC density functionals have been proposed
using the KED.\cite{2002BH,2002EMS,2002MES}
%--------------------------

\section{\label{sec:Functionals}Simple kinetic energy density functionals}

\subsection{\label{sec:first}The First Functional Approximations.}

The first explicit kinetic energy density functional formulated was the
Thomas-Fermi (TF) approximation,\cite{1927Th,1927Fer} where each point of an
inhomogeneous electron system with density $n(\mathbf{r})$ contributes to the
kinetic energy as any point in an homogeneous one with the same electron
density, $n_{0}=n(\mathbf{r})$. 
We then get the TF functional
\begin{equation}
T_{TF}[n(\mathbf{r})]=\int d\mathbf{r}C_{TF}n^{5/3}(\mathbf{r}), \label{eq:TF}%
\end{equation}
where $C_{TF}=\frac{3}{10}(3\pi^{2})^{2/3}$ is the TF constant. By
construction, this functional is exact for homogeneous systems and a good
approximation for systems close to the free electron
gas.\cite{b1983LundMarch,r1981Lieb} But this functional gives, in general,
poor results when applied to atoms or molecules, and the density profiles
obtained in a variational minimization show no quantum effects. For atoms, in
fact, no shell structure is obtained and the errors in the total kinetic
energy are about 10\%. This relative error is also obtained when the
functional is applied using \emph{good} densities.

Another explicit kinetic energy density functional is the von Weizs\"{a}cker
(vW) $T_{vW}[n]$\ functional, constructed to be exact for one or two electrons
in the same spacial state.\cite{1935Weiz} The most usual form in which
functional is found in the literature is
\begin{equation}
T_{vW}[n] = \frac{1}{8}\int{\frac{|\mathbf{\nabla} n(\mathbf{r})|^{2}%
}{n(\mathbf{r})}d\mathbf{r}}, \label{eq:vW}%
\end{equation}
but can also rewritten in other ways that yield the same total kinetic energy.
Being the functional exact for single orbital systems, it gives the correct
KED when the contribution of a given orbital to the electron density is much
larger than all the other orbitals. For the same reason, for those regions in
space that are very close or very far away from the positions of the nuclei,
this functional also gives the correct KED. Finally, $T_{vW}[n]$ is exact in
regions with large variations of the electron density. But when applied to
general systems this functional yields large errors. So, the variational
minimization for atoms gives no shell structures and the relative errors
increase a lot when the number of electrons grows, getting energies of a
different order of magnitude than the exact ones for atoms with a large number
of electrons.

A very interesting functional can be obtained as a combination of the TF 
and vW
functionals, the so called second order gradient expansion approximation
(GEA2), constructed as:\cite{1957Kirzhnits}
\begin{align}
T_{S}^{GEA2}[n]  &  =T_{TF}[n]+\frac{1}{9}T_{vW}[n]\nonumber\\
&  = C_{TF}\int{n(\mathbf{r})^{5/3}d\mathbf{r}} + \frac{1}{72}\int
{\frac{|\mathbf{\nabla}n(\mathbf{r})|^{2}} {n(\mathbf{r})}d\mathbf{r}}\\
&  = C_{TF}\int{d\mathbf{r}n(\mathbf{r})^{5/3}}\left[  1+\frac{1}{72C_{TF}%
}\left\vert \frac{\mathbf{\nabla}n(\mathbf{r})}{n^{4/3}(\mathbf{r}%
)}\right\vert ^{2}\right]  . \nonumber\label{eq:GEA2}%
\end{align}
This functional gives the correct kinetic energies for systems with slow
varying electron densities. Density profiles obtained in the variational
minimization of the total energy with the GEA2 approximation show the same
pathologies as the previous functionals, but when applied using \emph{good}
electron densities the error in the total energy is only about 1\% for a
number of very different systems. This error is clearly too big for chemical
precision, but is quite small for common orbital-free kinetic energy density
functionals.\cite{2001IEMS}

%--------------------------

\subsection{\label{sec:GGA}The Generalized Gradient Approximations.}

In this paper we make a comprehensive study of the kinetic energy density
functionals that can be expressed within the Generalized Gradient
Approximation (GGA), which allows a general form for those semilocal
functionals that only depend on the electron density and its gradient. Any
semilocal functional of this kind can be written as
\begin{equation}
T_{S}^{GGA}[n(\mathbf{r})]=\int d\mathbf{r}t_{TF}\left(  \mathbf{r}\right)
F_{enh}(s(\mathbf{r})),
\end{equation}
where $t_{TF}\left(  \mathbf{r}\right)  = C_{TF} n^{5/3}(\mathbf{r})$ is the
KED corresponding to the TF functional and $F_{enh}(s(\mathbf{r}))$ is the
so-called \emph{enhancement factor}, that depends on the adimensional
variable
\begin{equation}
s(\mathbf{r}) = \frac{\left\vert \mathbf{\nabla}n(\mathbf{r})\right\vert
}{n^{4/3}(\mathbf{r})}.
\end{equation}
The quantity $s(\mathbf{r})$ is called the \emph{reduced density gradient} and
has a clear physical interpretation, because it controls the speed of the
variation of the electron density. Large values of $s(\mathbf{r})$\ correspond
to fast variations in the electron density and small values to slow ones; a
zero value indicates a region of the space where the electron density has no variation.

The mathematical form of the \emph{enhancement factor}\ determines the
functional and authors have proposed many different forms for the enhancement
factor. In this paper, the following set of semilocal functionals have been selected.

\vspace{1cm}

\textbf{1. Thomas--Fermi functional (TF).}\cite{1927Th,1927Fer} When the
\emph{enhancement factor} has a constant value of $1$, the TF functional is
recovered,
\[
F_{TF}(s(\mathbf{r})) = 1.
\]

\textbf{2. Second order gradient expansion approximation (GEA2).}%
\cite{1957Kirzhnits} This approximation is a particular case of the GGA
approximations, obtained with the expression:
\[
F_{GEA2}(s(\mathbf{r})) = \left[  1 + \frac{1}{72C_{TF}} s(\mathbf{r}%
)^{2}\right]  .
\]

There are some functionals constructed as a linear combination of the TF
functional and the vW functional. We cite now some of them.

\textbf{3. Thomas--Fermi +} $\frac{1}{5}$ \textbf{von Weizs\"{a}cker
(TF}$\mathbf{5}$\textbf{W).} In this case the contribution of the vW
functional is weighted by a prefactor $1/5$,
\cite{1965YoneiTomish,1966TomishYonei,1967Yonei,1982Yonei,%
1979GrossDreiz,1983Berk,1982Stich,1994GWG,2001CCH}
\[
F_{TF5W}(s(\mathbf{r}))=\left[  1+\frac{1}{40C_{TF}}s(\mathbf{r}%
)^{2}\right]  .
\]

\textbf{4. Thomas--Fermi + von Weizs\"{a}cker (TFvW).}\cite{1967Yonei} The
combination involving both the full Thomas-Fermi and von Weizs\"{a}cker
functionals is theoretically interesting but in practical applications largely
overestimate the kinetic energy,
\[
F_{TFvW}(s(\mathbf{r}))=\left[  1+\frac{1}{8C_{TF}}s(\mathbf{r})^{2}\right]
.
\]

\textbf{5. Thomas--Fermi + } $\frac{b}{9}$ \textbf{von Weizs\"{a}cker
(TF}$\mathbf{9}$\textbf{W).} A modification of GEA2 can be written with the
introduction of a parameter $b$ that modifies the contribution of the gradient
correction. Several values of $b$ have been used by different authors. For this
work we have chosen a value of $b=1.067$.\cite{1992Thakk}
\[
F_{TF9W}(s(\mathbf{r}))=\left[  1+\frac{b}{72C_{TF}}s(\mathbf{r}%
)^{2}\right]  .
\]

\textbf{6. $N$-dependent Thomas--Fermi functional (TF-}$\mathbf{N}$%
\textbf{)}.\cite{1990ThakPed} Several functionals have been developed as
modifications of the TF functional by including a prefactor that depends on
the number of electrons $N$. For this work we have chosen:
\[
F_{TF-N}(s(\mathbf{r}))=\left(  1+\frac{0.313}{N^{1/3}}-\frac{0.187}{N^{2/3}%
}\right)  ,
\]
that usually provides better values for the kinetic energies than those
obtained with the TF functional.

\textbf{7. Pearson functional (Pear).}\cite{1982Pearson} This functional, a
modification of the GEA2 approximation that follows the idea presented by
Pearson and Gordon,\cite{1985Pearson} is constructed in such a way that the
gradient correction takes only into account the regions of the space where the
density varies slowly. This can be done with the introduction of a cut-off
that depends on the value of the reduced gradient $s$. A sharp cutoff usually
introduces numerical problems and a functional with a smooth
cutoff\cite{1982Pearson} is proposed with the \emph{enhancement factor}:
\[
F_{Pear}[n]=\int{d\mathbf{r}}\left[  1+\frac{1}{1+\left[  s/\zeta\right]
^{6}}\frac{1}{72C_{TF}}s(\mathbf{r})^{2}\right]  ,
\]
where $\zeta$ is a parameter that Pearson and Gordon fixed as $\zeta=1$, the
optimum value for fitting to atomic Hartree-Fock kinetic energies. This
functional form prevents the contributions of the gradient expansions coming
from regions where fast variations of the electron density occur.

\textbf{8. DePristo--Kress functional (DK).}\cite{1987DK} These authors
introduce a mathematical form based on a Pad\'{e}-type approximation that is
able to recover different desirable limits:%

\[
F_{DK}(s(\mathbf{r})) = \frac{9b_{3}x^{4}+a_{3}x^{3}+a_{2}x^{2}+a_{1}%
x+1}{b_{3}x^{3}+b_{2}x^{2}+b_{1}x+1},
\]
where
\[
x=\frac{(s(\mathbf{r}))^{2}}{72C_{TF}}.
\]
This functional is equivalent to GEA2 for slowly varying densities -- small
values of $s$ -- and equivalent to the von Weizs\"{a}cker functional for fast
varying densities -- large values of $s$ --. The parameters of the functional
($a_{1} = 0.95$, $a_{2} = 14.28111$, $a_{3} = -19.5762$, $b_{1} =-0.05$,
$b_{2} = 9.99802$, $b_{3} = 2.96085$) were obtained by fitting the results of
the functional to atomic Hartree-Fock kinetic energies:

\textbf{9. Lee--Lee--Parr functional (LLP).}\cite{1991LLP} These authors have
used the same mathematical form of an \emph{enhancement factor} for the
exchange functional for constructing the kinetic energy density functional
\[
F_{LLP}(s(\mathbf{r}))=\left[  1+\frac{b\left[  s(\mathbf{r})\right]  ^{2}%
}{1+c\left[  s(\mathbf{r})\right]  \arcsin(s(\mathbf{r}))}\right]  ,
\]
where the parameters have the values $b=0.0044188$ and $c=0.0253$.

\textbf{10. Ou-Yang -- Levy 1 functional (OL1).}\cite{1991OL} Following the
nonuniform coordinate scaling requirement for the kinetic energy density
functional,\cite{1990OLnonunif} these authors suggest these two new functional
forms:
\begin{equation}
F_{OL1}(s(\mathbf{r}))=1+\frac{1}{72C_{TF}}\left[  s(\mathbf{r})\right]
^{2}+d\left[  s(\mathbf{r})\right]  ,
\end{equation}
where $d=0.00187$, and

\textbf{11. Ou-Yang -- Levy 2 functional (OL2).}\cite{1991OL}%

\[
F_{OL2}(s(\mathbf{r})) = 1+\frac{1}{72C_{TF}}\left[  s(\mathbf{r})\right]
^{2}+\frac{D\left[  s(\mathbf{r})\right]  }{1+2^{5/3}\left[  s(\mathbf{r}%
)\right]  },
\]
with $D=0.0245$.

\textbf{12. Thakkar functional (Thak).}\cite{1992Thakk} 
In a review of many semilocal functionals, 
Thakkar proposed a new one as a conjoint of the
mathematical forms of the most successful functionals:
\[
F_{Thak}(s(\mathbf{r}))=1+\frac{b\left[  s(\mathbf{r})\right]  ^{2}%
}{1+c\left[  s(\mathbf{r})\right]  \arcsin[s(\mathbf{r})]}-\frac{D\left[
s(\mathbf{r})\right]  }{1+2^{5/3}\left[  s(\mathbf{r})\right]  },
\]
with the parameters: $b=0.0055$, $c=0.0253$ and $D=0.072$.

As was commented before, Lee, Lee and Parr argued that the 
enhancement factors used for the GGA XC energy functionals 
can also be used for the development of
kinetic energy functionals. The next five functionals, 
formulated following
that approach, were proposed by Lacks and Gordon.\cite{1994LG}

\textbf{13. Becke 86A functional (B86A).}\cite{1986aBecke}%

\[
F_{B86A}(s(\mathbf{r}))=1+0.0039\frac{\left[  s(\mathbf{r})\right]  ^{2}%
}{1+0.004\left[  s(\mathbf{r})\right]  ^{2}}.
\]

\textbf{14. Becke 86B functional (B86B).}\cite{1986bBecke}
\[
F_{B86B}(s(\mathbf{r}))=1+0.00403\frac{\left[  s(\mathbf{r})\right]  ^{2}%
}{\left(  1+0.007\left[  s(\mathbf{r})\right]  ^{2}\right)  ^{4/5}}.
\]

\textbf{15. DePristo--Kress 87 functional (DK87).}\cite{1987DKxc}
\begin{eqnarray*}
\lefteqn{%
F_{DK87}(s(\mathbf{r}))=1+ }  \\
& &
\frac{7}{324\left(  18\pi^{4}\right)  ^{1/3}}\left[
s(\mathbf{r})\right]  ^{2}\frac{1+0.861504s(\mathbf{r})}{1+0.044286\left[
s(\mathbf{r})\right]  ^{2}}.
\end{eqnarray*}

\textbf{16. Perdew--Wang 86 functional (PW86).}\cite{1986PWggaxc}
\[
F_{PW86}(s(\mathbf{r}))=1+1.296\left[  s(\mathbf{r})\right]  ^{2}+14\left[
s(\mathbf{r})\right]  ^{4}+0.2\left[  s(\mathbf{r})\right]  ^{61/15}.
\]

\textbf{17. Perdew--Wang 91 functional (PW91).}\cite{1992PerdewWAng}
\begin{eqnarray*}
\lefteqn{%
F_{PW91}(s(\mathbf{r}))= }   \\
& &
\frac{1+a_{1}s(\mathbf{r})\arg\sinh\left(  b\left[
s(\mathbf{r})\right]  \right)  +\left[  a_{2}-a_{3}e^{-100\left[
s(\mathbf{r})\right]  ^{2}}\right]  }%
{1+as(\mathbf{r})\arg\sinh\left(
b\left[  s(\mathbf{r})\right]  \right)  +a_{4}\left[  s(\mathbf{r})\right]
^{2}}.
\end{eqnarray*}
with $a_{1}=0.19645$, $a_{2}=0.2747$, $a_{3}=0.1508$, $a_{4}=0.004$ and
$b=7.7956$.

\textbf{18. Lacks--Gordon functional (LG94).} 
In their extensive study of the
kinetic energy density functionals,\cite{1994LG} these authors 
made their own
contribution with the formulation of a new one, based on a 
previous exchange
functional of the same authors developed by fitting exchange 
energies of atoms
and ions, as well as the correct behavior of the exchange 
energy for small $s(r)$:%

\begin{widetext}
\[
F_{LG94}(s(\mathbf{r})) = \left[  \frac{1+a_{2}\left[  s(\mathbf{r})\right]
^{2}+a_{4}\left[  s(\mathbf{r})\right]  ^{4}+a_{6}\left[  s(\mathbf{r}%
)\right]  ^{6}+a_{8}\left[  s(\mathbf{r})\right]  ^{8}+a_{10}\left[
s(\mathbf{r})\right]  ^{10}+a_{12}\left[  s(\mathbf{r})\right]  ^{12}%
}{1+10^{-8}\left[  s(\mathbf{r})\right]  ^{2}}\right]  ^{b},
\]
\end{widetext}
with $a_{2} =\left(  10^{-8}+0.1234\right)  /0.024974$, $a_{4} = 29.790$,
$a_{6} = 22.417$, $a_{8} = 12.119$, $a_{10} = 1570.1$, $a_{12} = 55.944$ and
$b = 0.024974$.

\textbf{19. von Weizs\"{a}cker functional (vW).}\cite{1935Weiz} 
The vW functional can also be written as a semilocal functional 
if we choose the
\emph{enhancement factor}:
\[
F_{vW}(s(\mathbf{r}))=\frac{1}{8C_{TF}}s(\mathbf{r})^{2}.
\]

\textbf{20. Acharya et al. functional (ABSP).}\cite{1980Acharya} Acharya,
Bartolotti, Sears, and Parr proposed a functional with the 
full vW functional,
introducing a TF contribution weighted by a prefactor depending 
on the electron number $N$:
\[
F_{ABSP}[n]=\frac{1}{8C_{TF}}s(\mathbf{r})^{2}+\left(  1-\frac{1.412}{N^{1/3}%
}\right)  .
\]

\textbf{21. G\'{a}zquez--Robles functional (GR).}\cite{1982GazRob} 
Following the same spirit of the previous functional, G\'{a}zquez 
and Robles developed another one, with a more complicated form 
for the weight of the TF functional:
\begin{equation}
F_{GR}[n]=\frac{1}{8C_{TF}}s(\mathbf{r})^{2}+\left(  1-\frac{2}{N}\right)
\left(  1-\frac{1.303}{N^{1/3}}+\frac{0.029}{N^{2/3}}\right)  .
\end{equation}

\section{\label{sec:measure}Measurement of the Distance Between 
two Kinetic Energy Densities.}

As the KED can be evaluated by using the KS orbitals, we may 
think that every approximate functional yields an approximation 
to the KED by simply defining
it as the integrand 
$t_{S}^{func}(\mathbf{r})$\ 
that appears in the expression of the functional:
\begin{equation}
T_{S}^{func}[n]=\int d\mathbf{r~}t_{S}^{func}(\mathbf{r}).
\end{equation}

As the main aim of this paper is to know, in a 
{\bf quantitative} way, 
how the semilocal functionals are able to reproduce the KED, 
we need to define a measure of the closeness between the 
orbital-based and approximated
\emph{orbital-free} KED distributions, i.e. we must find 
an efficient and reasonable measure of the distances 
between distributions. 
Some distances between probability distributions have been 
defined, most of them to be used for normalized and positive 
definite distributions. 
The electron density is normalized and positive, but the 
KEDs are no normalized and they are not, in general, positive 
definite (note that the second definition of the KS KED has
always negative values in some regions of the space). 
For these reasons those distances are not appropriate in 
our case. But the absolute value norm, being
the absolute value of the difference between two 
distributions integrated over the whole space, 
is a measure of the accumulated local error of the KED,
\begin{equation}
\epsilon=\int d\mathbf{r}\left\vert t_{S}(\mathbf{r})-t_{S}^{func}%
(\mathbf{r})\right\vert ,
\end{equation}
where $t_{S}(\mathbf{r})$ is a valid distribution of KED and 
$t_{S}^{func}(\mathbf{r})$\ an approximated one. 
This definition satisfies the three
desirable requirements for a distance: 
it is positive definite, is symmetric
and verifies the triangular inequality.

The defined measure $\epsilon$ depends on the size of 
the system, because the
total kinetic energy has a nonlinear dependence with 
the number of electrons
(the kinetic energy of neon is more than two hundred 
times the kinetic energy
of hydrogen). 
In order to get a measure that does not depend so much 
on the size of the system we divide $\epsilon$ by the 
exact total kinetic energy. 
In this way, we propose a \emph{quality factor} 
$\sigma$\ 
for a KED as
\begin{equation}
\sigma=\frac{\epsilon}{T_{S}[n]}.
\end{equation}
With this definition, the bigger the difference 
between the distributions is
the larger the value of $\sigma$ becomes, and $\sigma$ 
is zero for two identical KEDs. 
Note that the value of $\sigma$ can be interpreted 
as the amount of the approximated KED that is misplaced 
when compared to the distribution of the orbital-based KED. 
With this factor $\sigma$ we are going
to test whether the good results obtained for the 
total kinetic energies with
the GGA functionals are due to their correct local KED 
or to error cancellations among different regions of 
the space.

%------------------------------------------------------------------------

\section{\label{sec:atoms}Electron Densities for Light Atoms}

In order to clarify the quality of the KED each approximate 
functional yields, 
we want to use good enough atomic electron densities to 
evaluate our semilocal functionals.

We initially tried commercial software that performs KS 
calculations and outputs the orbitals used; 
in that way the related electron density, the KED
and the total kinetic energy can be evaluated. 
But we decided to avoid the use
of these codes for several reasons. 
Firstly, the behavior of any XC functional
leads to a bad decay of the density in the outer region 
of the atom,
reflecting some problems in the quality of the KS orbitals, 
their density and in the KED. 
More importantly, when using a basis set of gaussians, several
gaussians are generally needed to approximate one Slater-like KS orbital. 
As a result, changes in the slope of the KS orbitals appear; 
that could not
noticeable by simple inspection but, when the careful 
numerical evaluation of the laplacian of the density is 
done, spurious structure arises. 
So, when the laplacian of the density is numerically 
evaluated one must be cautious with
the use of a gaussian basis set. 
Furthermore, gaussian basis set are unable to
give a good description of both the cusps of the density 
in the positions of
the nuclei and the decay of the orbitals far away from them.

For that reasons, we have decided to use Slater orbitals 
for the light atoms.
They only approximately describe the KS orbitals but 
all the aforementioned pathologies will not appear 
(there are no spurious oscillations, no cut-offs
and correct cusp conditions and decay of the orbitals 
can be achieved). 
There are several sets of values of exponents for the 
Slater orbitals, with only small differences between 
them; Table~\ref{Table:Zeff} presents our choice of
parameters\cite{b1997Atkins,1963CR} for the ten first 
atoms.

%---------------------------------------------------
%\input{Tabla_Zeff.tex}
\begin{table*}[t] 
\centering%

% Table generated by Excel2LaTeX from sheet 'LapOrb'
\begin{tabular}{||c||cccccccccc||}
\hline\hline
& { H} & { He} & { Li} & { Be} & 
{ B} & { C} & { N} & { O} & { F} & { Ne}
\\ \hline\hline
{ 1s} & { 1} & { 1.6875} & { 2.6906} & { 3.6848%
} & { 4.6795} & { 5.6727} & { 6.665} & { 7.6579} & 
{ 8.6501} & { 9.6421} \\ 
{ 2s} &  &  & { 1.2792} & { 1.912} & { 2.5762} & 
{ 3.2166} & { 3.8474} & { 4.4916} & { 5.1276} & 
{ 5.7584} \\ 
{ 2p} &  &  &  &  & { 2.4214} & { 3.1358} & { 3.834}
& { 4.4532} & { 5.1000} & { 5.7584} \\ \hline\hline
\end{tabular}
\caption{Values of the effective charge, $Z_{eff}$.}%
\label{Table:Zeff}%
\end{table*}
%---------------------------------------------------

%------------------------------------------------------------------------

\section{\label{sec:total}Total Kinetic Energies}

We have performed total energy calculations for twenty 
one functionals, using
the densities corresponding to the Slater orbitals 
with parameters given in Table~\ref{Table:Zeff}.

The relative errors for the total kinetic energies are 
presented in Table~\ref{Table:TsErSL}. 
We also show the average of the absolute values of
the errors for every functional. 
To add more information, the dispersion of
data is displayed under the label 
\textquotedblleft Range", where we give the
statistical range, i.e., the length of the interval 
which includes all the former data. 
When the value of the range is larger than the average 
we can think the average could be meaningless.
All the tables of this paper will be presented this way 
(in each case, for the relative error, for the values 
of $\sigma$, etc.).

%----------------------------------------------------------
%\input{Tabla_ErrRelativo.tex}
\begin{table*}[t] 
\centering%

%Tabla_ErrRelativo.tex

% Table generated by Excel2LaTeX from sheet 'FormattedTable'
\begin{tabular}{||c||cccccccccc||c||c||}
\hline\hline
       & { H} & { He} & { Li} & { Be} & { B}
& { C} & { N} & { O} & { F} & { Ne} & { Ave} 
& { Range} 
\\ \hline\hline
{ TF} & { -0.082} & { -0.082} & { -0.085} & { %
-0.091} & { -0.097} & { -0.095} & { -0.085} & { %
-0.090} & { -0.087} & { -0.077} & { 0.087} & { 0.020}
\\ 
{ GEA2} & { 0.029} & { 0.029} & { 0.023} & { 0.013%
} & { 0.003} & { 0.000} & { 0.006} & { -0.003} & 
{ -0.003} & { 0.004} & { 0.011} & { 0.032}
\\ 
{ TF5W} & { 0.118} & { 0.118} & { 0.109} & { 0.097%
} & { 0.083} & { 0.077} & { 0.079} & { 0.067} & 
{ 0.064} & { 0.068} & { 0.088} & { 0.054}
\\ 
{ TFvW} & { 0.918} & { 0.918} & { 0.887} & { 0.846%
} & { 0.800} & { 0.762} & { 0.736} & { 0.695} & 
{ 0.667} & { 0.648} & { 0.788} & { 0.269}
\\ 
{ TF9W} & { 0.036} & { 0.036} & { 0.030} & { 0.020%
} & { 0.010} & { 0.007} & { 0.013} & { 0.003} & 
{ 0.003} & { 0.009} & { 0.017} & { 0.034}
\\ 
{ TF-N} & { 0.034} & { 0.038} & { 0.031} & { 0.021%
} & { 0.011} & { 0.010} & { 0.018} & { 0.010} & 
{ 0.011} & { 0.020} & { 0.020} & { 0.028}
\\ 
{ Pear} & { -0.070} & { -0.070} & { -0.074} & { %
-0.079} & { -0.085} & { -0.083} & { -0.073} & { %
-0.078} & { -0.075} & { -0.064} & { 0.075} & { 0.021}
\\ 
{ DK} & { 0.031} & { 0.031} & { 0.016} & { %
-0.003} & { -0.011} & { -0.010} & { -0.001} & { %
-0.007} & { -0.005} & { 0.005} & { 0.012} & { 0.042}
\\ 
{ LLP} & { 0.023} & { 0.023} & { 0.018} & { 0.011%
} & { 0.002} & { 0.001} & { 0.008} & { 0.000} & 
{ 0.001} & { 0.008} & { 0.010} & { 0.022}
\\ 
{ OL1} & { 0.039} & { 0.039} & { 0.032} & { %
0.023} & { 0.012} & { 0.010} & { 0.016} & { 0.006} & 
{ 0.006} & { 0.013} & { 0.020} & { 0.033}
\\ 
{ OL2} & { 0.036} & { 0.036} & { 0.029} & { %
0.020} & { 0.009} & { 0.007} & { 0.013} & { 0.004} & 
{ 0.004} & { 0.011} & { 0.017} & { 0.032}
\\ 
{ Thak} & { 0.026} & { 0.026} & { 0.022} & { %
0.015} & { 0.006} & { 0.005} & { 0.011} & { 0.004} & 
{ 0.004} & { 0.011} & { 0.013} & { 0.022}
\\ 
{ B86A} & { 0.023} & { 0.023} & { 0.019} & { %
0.013} & { 0.004} & { 0.003} & { 0.010} & { 0.002} & 
{ 0.003} & { 0.010} & { 0.011} & { 0.021}
\\ 
{ B86B} & { 0.049} & { 0.049} & { 0.045} & { %
0.038} & { 0.028} & { 0.026} & { 0.033} & { 0.025} & 
{ 0.025} & { 0.032} & { 0.035} & { 0.025}
\\ 
{ DK87} & { 0.030} & { 0.030} & { 0.026} & { %
0.019} & { 0.010} & { 0.009} & { 0.015} & { 0.008} & 
{ 0.008} & { 0.015} & { 0.017} & { 0.022}
\\ 
{ PW86} & { 0.020} & { 0.020} & { 0.016} & { %
0.009} & { 0.001} & { 0.000} & { 0.007} & { -0.001}
& { 0.000} & { 0.008} & { 0.008} & { 0.021} 
\\ 
{ PW91} & { 0.026} & { 0.026} & { 0.021} & { %
0.014} & { 0.005} & { 0.003} & { 0.009} & { 0.001} & 
{ 0.001} & { 0.008} & { 0.011} & { 0.025}
\\ 
{ LG94} & { 0.024} & { 0.024} & { 0.020} & { %
0.013} & { 0.005} & { 0.004} & { 0.010} & { 0.003} & 
{ 0.003} & { 0.011} & { 0.012} & { 0.021}
\\ 
{ vW} & { 0.000} & { 0.000} & { -0.027} & { %
-0.064} & { -0.104} & { -0.143} & { -0.179} & { %
-0.215} & { -0.247} & { -0.275} & { 0.125} & { 0.275}
\\ 
{ ABSP} & { -0.378} & { -0.111} & { -0.008} & { %
0.037} & { 0.054} & { 0.059} & { 0.061} & { 0.052} & 
{ 0.047} & { 0.043} & { 0.085} & { 0.439}
\\ 
{ GR} & { 0.252} & { 0.000} & { 0.006} & { %
0.023} & { 0.031} & { 0.033} & { 0.035} & { 0.028} & 
{ 0.023} & { 0.022} & { 0.045} & { 0.252} 
\\ 
\hline\hline
\end{tabular}
\caption{Relative errors in the total kinetic energy for the 
semilocal functionals. 
The average is made over the absolute relative errors.}%
\label{Table:TsErSL}%

\end{table*}
%----------------------------------------------------------

As we can see in Table~\ref{Table:TsErSL}, we get errors 
bigger than $8\%$ for
the $T_{TF}$, $T_{TF1/5vW}$, $T_{TFvW}$, $T_{Pear}$, 
$T_{vW}$, $T_{ABSP}$
functionals, about $4.5\%$ for the $T_{GR}$ one and 
smaller than $2\%$ for all
the other functionals.

%------------------------------------------------------------------------

\section{\label{sec:laplacian}The Kinetic Energy Density and 
the Role of the
Laplacian of the Electron Density}

The nonuniqueness of the orbital-based definition of the KED 
have been pointed out by many authors. 
The first definition $t_{S}^{I}(\mathbf{r})$ is widely
used in the theory \emph{Atoms in Molecules}\cite{b1990Bader} 
of Bader, that
recently claimed the usefulness of the topological analysis 
of that definition\cite{b2005Bader}. 
The asymptotic behavior of the of the KED seems to support 
the use of the second definition 
$t_{S}^{II}\left(  \mathbf{r}\right)  $.\cite{1996YLW} 
But \textquotedblleft classical" properties are
recovered more appropriately with a mixture of both 
definitions, in particular the mean average of them 
(see, e. g., Refs. \onlinecite{1984GBP} and
\onlinecite{1991LMM}). 
We feel the use of one or another definition is still
an open question. 

When we compare the approximated orbital-free KED\ with
$t_{S}^{I}(\mathbf{r})$ very large values for the 
quality factor $\sigma$ are
obtained, about 0.6 for the TF functional and for 
almost all the other GGA functionals.
But, as commented, the comparison of the GGA functionals 
with any single
definition of KED is only a choice. 
It is possible to make comparisons with an
infinite set of KEDs. Indeed, the difference 
between $t_{S}^{I}(\mathbf{r})$
and $t_{S}^{II}(\mathbf{r})$ is one fourth the laplacian of 
the electron
density, 
an archetypical function related with the electron system 
that
integrates to zero over the whole space and having the 
adequate scaling properties to be a kinetic energy density. 
So, an infinite set of valid KEDs
can be obtained through (see, e.g., Ref. \onlinecite{1996YLW})
\begin{equation}
t_{S}^{L}(\mathbf{r})=t_{S}^{I}(\mathbf{r})+at_{0}(\mathbf{r}%
),\label{eq:tslapden}%
\end{equation}
where the $t_{S}^{L}(\mathbf{r})$ is constructed as 
$t_{S}^{I}(\mathbf{r})$
plus the laplacian of the electron density 
$t_{0}(\mathbf{r})=\nabla^{2}n(\mathbf{r})$, 
multiplied by a prefactor $a$ that can have any real
value. 
The value of $a=0$ yields $t_{S}^{I}(\mathbf{r})$, 
whereas the value of
$a=-\frac{1}{4}$ gives $t_{S}^{II}(\mathbf{r})$. 
On the other hand, the
arithmetic mean of the first and second definition 
is recovered with $a=-\frac{1}{8}$,\cite{1984GBP,1991LMM} 
and we have all the intermediate KEDs by
continuously varying~$a$.

We want to test every functional in the more adequate 
conditions for itself. 
For that reason, we have compared its approximate 
KED with the $t_{S}^{L}(\mathbf{r})$ that is its closest, 
choosing among all possible values of
the parameter $a$. 
To do that we minimize the value of $\sigma$ when varying
the parameter $a$. 
To obtain the value of $a$ that makes 
the best fit of the
distribution $t_{S}^{L}(\mathbf{r})$ to the 
approximated KED we have minimized
the value of $\sigma$ using a \textit{golden search} 
algorithm.\cite{bNumRec}
After the minimization process, each functional has a 
given value of $a$ that
yields the lower value of $\sigma$, being the closest 
KED that 
$t_{S}^{L}(\mathbf{r})$ constructed by using this 
value of the parameter $a$ in
Eq.~(\ref{eq:tslapden}).

%----------------------------------------------------
%\input{Tabla_aLapDens.tex}
\begin{table*}[t] 
\centering%

%Tabla_aLapDens.tex
\begin{tabular}{||c||cccccccccc||c||c||}
\hline\hline
& { H} & { He} & { Li} & { Be} & { B} & 
{ C} & { N} & { O} & { F} & { Ne} & { %
Ave.} & { Range} \\ \hline\hline
{ TF} & { 0.163} & { 0.163} & { 0.166} & { 0.169%
} & { 0.169} & { 0.169} & { 0.169} & { 0.169} & 
{ 0.169} & { 0.168} & { 0.167} & { 0.006} \\ 
{ GEA2} & { 0.144} & { 0.144} & { 0.147} & { 0.149%
} & { 0.150} & { 0.151} & { 0.151} & { 0.152} & 
{ 0.152} & { 0.152} & { 0.149} & { 0.008} \\ 
{ TF5W} & { 0.129} & { 0.129} & { 0.132} & { 0.135%
} & { 0.136} & { 0.138} & { 0.138} & { 0.140} & 
{ 0.140} & { 0.140} & { 0.136} & { 0.011} \\ 
{ TFvW} & { 0.000} & { 0.000} & { 0.046} & { 0.066%
} & { 0.074} & { 0.080} & { 0.083} & { 0.089} & 
{ 0.095} & { 0.097} & { 0.063} & { 0.097} \\ 
{ TF9W} & { 0.143} & { 0.143} & { 0.146} & { 0.148%
} & { 0.149} & { 0.150} & { 0.150} & { 0.151} & 
{ 0.151} & { 0.151} & { 0.148} & { 0.008} \\ 
{ TF-N} & { 0.162} & { 0.162} & { 0.165} & { 0.167%
} & { 0.167} & { 0.168} & { 0.167} & { 0.167} & 
{ 0.167} & { 0.167} & { 0.166} & { 0.006} \\ 
{ Pear} & { 0.164} & { 0.164} & { 0.168} & { 0.170%
} & { 0.171} & { 0.171} & { 0.171} & { 0.170} & 
{ 0.170} & { 0.170} & { 0.169} & { 0.007} \\ 
{ DK} & { 0.144} & { 0.144} & { 0.149} & { 0.160%
} & { 0.159} & { 0.159} & { 0.158} & { 0.158} & 
{ 0.157} & { 0.157} & { 0.155} & { 0.016} \\ 
{ LLP} & { 0.150} & { 0.150} & { 0.151} & { 0.153%
} & { 0.154} & { 0.155} & { 0.155} & { 0.155} & 
{ 0.155} & { 0.155} & { 0.153} & { 0.006} \\ 
{ OL1} & { 0.143} & { 0.143} & { 0.146} & { %
0.148} & { 0.149} & { 0.150} & { 0.151} & { 0.151} & 
{ 0.151} & { 0.151} & { 0.148} & { 0.008} \\ 
{ OL2} & { 0.144} & { 0.144} & { 0.147} & { %
0.149} & { 0.150} & { 0.151} & { 0.151} & { 0.152} & 
{ 0.152} & { 0.152} & { 0.149} & { 0.008} \\ 
{ Thak} & { 0.148} & { 0.148} & { 0.150} & { %
0.152} & { 0.153} & { 0.153} & { 0.153} & { 0.153} & 
{ 0.154} & { 0.153} & { 0.152} & { 0.005} \\ 
{ B86A} & { 0.149} & { 0.149} & { 0.151} & { %
0.153} & { 0.154} & { 0.154} & { 0.154} & { 0.154} & 
{ 0.155} & { 0.154} & { 0.153} & { 0.005} \\ 
{ B86B} & { 0.147} & { 0.147} & { 0.148} & { %
0.151} & { 0.151} & { 0.152} & { 0.152} & { 0.152} & 
{ 0.152} & { 0.152} & { 0.150} & { 0.005} \\ 
{ DK87} & { 0.148} & { 0.148} & { 0.150} & { %
0.152} & { 0.152} & { 0.153} & { 0.153} & { 0.152} & 
{ 0.153} & { 0.152} & { 0.151} & { 0.005} \\ 
{ PW86} & { 0.150} & { 0.150} & { 0.152} & { %
0.154} & { 0.155} & { 0.155} & { 0.155} & { 0.155} & 
{ 0.155} & { 0.156} & { 0.154} & { 0.005} \\ 
{ PW91} & { 0.147} & { 0.147} & { 0.148} & { %
0.151} & { 0.152} & { 0.152} & { 0.152} & { 0.152} & 
{ 0.153} & { 0.152} & { 0.151} & { 0.006} \\ 
{ LG94} & { 0.148} & { 0.148} & { 0.149} & { %
0.151} & { 0.152} & { 0.152} & { 0.152} & { 0.152} & 
{ 0.152} & { 0.152} & { 0.151} & { 0.005} \\ 
{ vW} & { 0.000} & { 0.000} & { 0.000} & { %
0.000} & { 0.002} & { 0.002} & { 0.002} & { 0.003} & 
{ 0.003} & { 0.003} & { 0.002} & { 0.003} \\ 
{ ABSP} & { 0.000} & { 0.000} & { 0.004} & { %
0.016} & { 0.025} & { 0.033} & { 0.040} & { 0.045} & 
{ 0.049} & { 0.053} & { 0.026} & { 0.054} \\ 
{ GR} & { 0.000} & { 0.000} & { 0.006} & { %
0.015} & { 0.023} & { 0.031} & { 0.037} & { 0.043} & 
{ 0.048} & { 0.051} & { 0.025} & { 0.051} \\ 
\hline\hline
\end{tabular}

\caption{Values of $a$ when the semilocal functionals are fitted 
to $t_{S}^{L}(\mathbf{r})$, 
Eq.~(\ref{eq:tslapden}).}%
\label{Table:aLapDens}%

\end{table*}
%----------------------------------------------------

This methodology is used for all the functionals. 
The best values of $a$ are
shown in Table~\ref{Table:aLapDens}. 
We see that the values of $a$ are almost
constant and do not depend so much 
on the number of electrons 
for almost all of the functionals. 
For all functionals but those with a full vW term, 
the best fits to $t_{S}^{L}(\mathbf{r})$ are almost 
equidistant to both $t_{S}^{I}(\mathbf{r})$ 
and $t_{S}^{II}(\mathbf{r})$, although a little 
bit closer to the second one. 
The vW functional is a special case: 
the values of $a$ are zero for the four first atoms 
and very close to zero for all the others, 
reflecting the fact that $T_{vW}$ 
in the form usually found in literature 
is directly related to 
$t_{S}^{I}(\mathbf{r})$ 
and its approximate KED is always closer to that 
definition than to any other one.

%--------------------------------------------------------
%\input{Tabla_SigLapDens.tex}
\begin{table*}[t] 
\centering%

\begin{tabular}{||c||cccccccccc||c||c||}
\hline\hline
& { H} & { He} & { Li} & { Be} & { B}
& { C} & { N} & { O} & { F} & { Ne} & { %
Ave.} & { Range}
\\ \hline\hline
{ TF} & { 0.166} & { 0.166} & { 0.166} & { 0.168%
} & { 0.171} & { 0.167} & { 0.162} & { 0.161} & 
{ 0.160} & { 0.160} & { 0.165} & { 0.012}
 \\ 
{ GEA2} & { 0.187} & { 0.187} & { 0.188} & { 0.190%
} & { 0.192} & { 0.188} & { 0.187} & { 0.186} & 
{ 0.185} & { 0.186} & { 0.188} & { 0.007}
\\ 
{ TF5W} & { 0.225} & { 0.225} & { 0.227} & { 0.228%
} & { 0.228} & { 0.225} & { 0.224} & { 0.221} & 
{ 0.220} & { 0.221} & { 0.224} & { 0.009} 
\\ 
{ TFvW} & { 0.918} & { 0.918} & { 0.893} & { 0.873%
} & { 0.838} & { 0.805} & { 0.779} & { 0.750} & 
{ 0.727} & { 0.710} & { 0.821} & { 0.208} 
\\ 
{ TF9W} & { 0.189} & { 0.189} & { 0.191} & { 0.193%
} & { 0.195} & { 0.190} & { 0.190} & { 0.188} & 
{ 0.187} & { 0.188} & { 0.190} & { 0.007}
\\ 
{ TF-N} & { 0.210} & { 0.213} & { 0.210} & { 0.207%
} & { 0.204} & { 0.197} & { 0.196} & { 0.191} & 
{ 0.190} & { 0.191} & { 0.201} & { 0.023}
\\ 
{ Pear} & { 0.166} & { 0.166} & { 0.166} & { 0.168%
} & { 0.170} & { 0.165} & { 0.162} & { 0.161} & 
{ 0.160} & { 0.161} & { 0.164} & { 0.010} 
\\ 
{ DK} & { 0.218} & { 0.218} & { 0.207} & { 0.208%
} & { 0.209} & { 0.203} & { 0.200} & { 0.197} & 
{ 0.194} & { 0.194} & { 0.205} & { 0.024}
\\ 
{ LLP} & { 0.199} & { 0.199} & { 0.199} & { 0.199%
} & { 0.200} & { 0.195} & { 0.193} & { 0.191} & 
{ 0.190} & { 0.191} & { 0.196} & { 0.010} 
\\ 
{ OL1} & { 0.192} & { 0.192} & { 0.193} & { %
0.195} & { 0.197} & { 0.192} & { 0.191} & { 0.190} & 
{ 0.189} & { 0.190} & { 0.192} & { 0.008} 
\\ 
{ OL2} & { 0.190} & { 0.190} & { 0.191} & { %
0.194} & { 0.195} & { 0.191} & { 0.190} & { 0.188} & 
{ 0.188} & { 0.189} & { 0.191} & { 0.008} 
\\ 
{ Thak} & { 0.203} & { 0.203} & { 0.202} & { %
0.202} & { 0.202} & { 0.197} & { 0.195} & { 0.193} & 
{ 0.191} & { 0.192} & { 0.198} & { 0.011}
\\ 
{ B86A} & { 0.203} & { 0.203} & { 0.202} & { %
0.202} & { 0.202} & { 0.196} & { 0.195} & { 0.192} & 
{ 0.191} & { 0.192} & { 0.198} & { 0.011} 
\\ 
{ B86B} & { 0.217} & { 0.217} & { 0.216} & { %
0.215} & { 0.215} & { 0.210} & { 0.208} & { 0.205} & 
{ 0.204} & { 0.204} & { 0.211} & { 0.013}
\\ 
{ DK87} & { 0.210} & { 0.210} & { 0.209} & { %
0.208} & { 0.208} & { 0.203} & { 0.201} & { 0.198} & 
{ 0.196} & { 0.197} & { 0.204} & { 0.013} 
\\ 
{ PW86} & { 0.201} & { 0.201} & { 0.200} & { %
0.200} & { 0.200} & { 0.195} & { 0.193} & { 0.191} & 
{ 0.190} & { 0.191} & { 0.196} & { 0.011} 
\\ 
{ PW91} & { 0.200} & { 0.200} & { 0.200} & { %
0.200} & { 0.201} & { 0.196} & { 0.194} & { 0.192} & 
{ 0.191} & { 0.191} & { 0.197} & { 0.010} 
\\ 
{ LG94} & { 0.207} & { 0.207} & { 0.207} & { %
0.206} & { 0.206} & { 0.200} & { 0.198} & { 0.195} & 
{ 0.193} & { 0.194} & { 0.201} & { 0.014} 
\\ 
{ vW} & { 0.000} & { 0.000} & { 0.027} & { %
0.064} & { 0.104} & { 0.143} & { 0.179} & { 0.215} & 
{ 0.247} & { 0.275} & { 0.125} & { 0.275}
\\ 
{ ABSP} & { 0.378} & { 0.111} & { 0.034} & { %
0.118} & { 0.173} & { 0.209} & { 0.232} & { 0.256} & 
{ 0.271} & { 0.280} & { 0.206} & { 0.345} 
\\ 
{ GR} & { 0.252} & { 0.000} & { 0.044} & { %
0.107} & { 0.156} & { 0.192} & { 0.217} & { 0.243} & 
{ 0.260} & { 0.270} & { 0.174} & { 0.270} 
\\ 
\hline\hline
\end{tabular}

\caption{Values of $\sigma$ when the semilocal functionals 
are fitted to $t_{S}^{L}(\mathbf{r})$, 
Eq.~(\ref{eq:tslapden}).}%
\label{Table:SigLapDens}%

\end{table*}
%--------------------------------------------------------

In Table~\ref{Table:SigLapDens} we present the corresponding 
values of
$\sigma$. 
As an average, the TF functional puts a 16.5\% of the KED 
misplaced from those regions where the KED that better 
fits TF is located. 
Thinking that the GGA functionals are corrections to the 
TF functional, one can expect that they will improve not 
only the total energies (as they do) but also the local
behavior of the kinetic functional. GEA2 do the same for 
18.8\%, and a careful
exam of the results show an unexpected result: 
all the GGA functionals,
despite their improvement in the evaluation of the total 
energies, yield larger values for the quality factor 
$\sigma$ that the TF ones. 
It seems that, within the GGA scheme, all functionals 
but those with a full vW term improves
the TF results giving total kinetic energies 
within 2\% of the exact one, but
they place the additional KED (i.e. the KED not 
included in the TF functional)
in wrong regions of the space. 
We conclude that the GGA functionals improve
the results for the TF kinetic energy by global error 
cancellations in the
evaluation of the total kinetic energies, while the 
local behavior of their
KEDs becomes worse than the TF one. The only exception 
to the previous results
is the Pearson functional, constructed with a different 
philosophy, which
gives slightly better values for $\sigma$ and for 
the total kinetic energies.

%------------------------------------------------------------------------

For the purpose of gaining insight on the origin of 
the previous results, we
have also divided the contribution to the KED coming from 
each orbital.
Even with no clear physical justification, we have used 
a sum of the laplacian for each
orbital $\phi_{i}(\mathbf{r})$\ with a parameter 
$a_{i}$ as a prefactor 
-- summation is extended over the $N$ electrons of the 
system. 
We assume that electrons with opposite spin, but 
sharing the same spatial orbitals, have the
same $a_{i}$. In the cases where $p$-orbitals are needed 
we only use one parameter for all them, in order to 
preserve the spherical symmetry. 
For our light atoms we then have 
$t_{S}^{L}(r)=t_{S}^{I}(r)+a_{1s}\nabla^{2}%
n_{1s}(r)+a_{2s}\nabla^{2}n_{2s}(r)+a_{2p}\nabla^{2}n_{2p}(r)$, 
where
$n_{1s}(r)$, $n_{2s}(r)$\ and $n_{2p}(r)$\ 
are the orbital densities, obtained
squaring the appropriate atomic orbitals. 
All quantities only depend on the
radial distance $r$. We have obtained the same 
qualitative behavior for the
values of $\sigma$ as those obtained with the previous 
method and we cannot
extract any additional information about the quality of 
the functionals in this way. 
As expected, we conclude that the use of the laplacian of 
the electron density is the correct way to generate a 
representative set of KEDs.

%------------------------------------------------------------------------

For the sake of completion, we also present the results 
obtained with a more
sophisticated fully nonlocal functional. 
By \textquotedblleft fully nonlocal"
we mean the functional explores the whole space when 
evaluating the contribution to the kinetic energy from 
any point of the system. 
We have chosen the functional developed in 1985 
by Chac\'{o}n-Alvarellos-Tarazona
\cite{1985CAT}, the simplest of a family of
functionals.\cite{1996GAC1,1996GAC2,1998GAC1,1998GAC2,2000GAC} 
In Table~\ref{Table:CAT} we show the relative errors 
for the total kinetic energies and the values of 
$\sigma$ obtained with the aforementioned procedure
and this functional approximation. Note the errors are 
about those obtained for most of the GGA functionals 
(in the average, smaller than $3.5\%$), 
and the values of $\sigma$ do not represent any clear 
improvement over the GGA ones. 
A complete study of the rest of the related fully 
nonlocal functionals, as well as a number of another 
kinetic energy functionals, has been done in
Ref. \onlinecite{2006ThesisDGA} and will be presented 
elsewhere.\cite{2007GAA2}

%------------------------------------------------------
%\input{Tabla_CAT.tex}
\begin{table*}[t] 
\centering%

%Tabla_CAT.tex

\begin{tabular}{||c||cccccccccc||c||c||}
\hline\hline
       & { H} & { He} & { Li} & { Be} & { B}
& { C} & { N} & { O} & { F} & { Ne} & { Ave} 
& { Range}  
\\ \hline\hline
{ Rel. error} & { 0.107} & { -0.041} & { -0.012} & { %
0.022} & { 0.021} & { 0.032} & { 0.037} & { %
0.032} & { 0.023} & { 0.011} & { 0.034} & { 0.148}
\\ 
{ $\sigma$} & { 0.107} & { 0.055} & { 0.082} & { 0.105%
} & { 0.122} & { 0.138} & { 0.155} & { 0.170} & 
{ 0.186} & { 0.205} & { 0.133} & { 0.150}
\\ 
\hline\hline
\end{tabular}
\caption{Relative errors in the total kinetic energy and 
values of $\sigma$ when adjusting with Eq.~(\ref{eq:tslapden})
for the original 1985 CAT functional. 
The average of the relative errors is made over their 
absolute values.}%
\label{Table:CAT}%

\end{table*}
%------------------------------------------------------

%------------------------------------------------------------------------

\section{A brief Graphical Study}

Up to now, integrated values for the study of the KED 
have been discussed. 
Now we present a graphical study of the KED for the neon 
atom; our aim is to show how the new technique developed 
works and the qualitative behavior of the approximate 
KEDs.

For the sake of briefness, we have chosen three 
representative functionals: 
the TF functional, the GEA2 approximation and the 
vW functional. 
TF yields the lowest values of the \emph{quality factor}, 
GEA2 is representative of those GGA functionals that give 
errors of about $1\%$ for the total kinetic energy and
the lowest values of $\sigma$ of the usual GGA functionals. 
Finally, vW is also studied due to its theoretical importance 
and its special behavior.

\begin{figure}[htbp]
\begin{center}
\includegraphics[height=3.5in,width=3.5in]%
{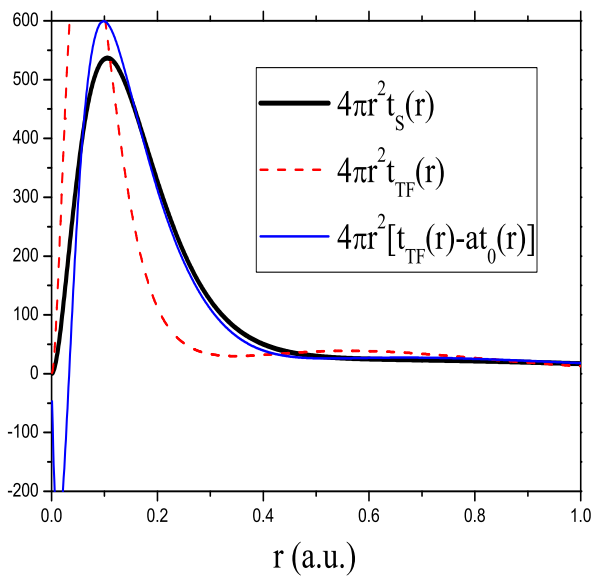}%
\caption{Radial TF kinetic energy densities for the neon atom.}%
\label{fig:NeTF}
\end{center}
\end{figure}

\begin{figure}[htbp]
\begin{center}
\includegraphics[height=3.5in,width=3.5in]%
{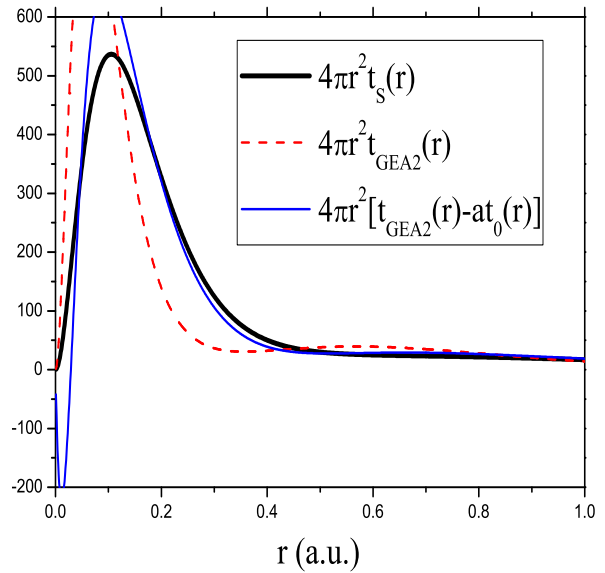}%
\caption{Radial GEA2 kinetic energy densities for the neon atom.}%
\label{fig:NeGEA}
\end{center}
\end{figure}

\begin{figure}[htbp]
\begin{center}
\includegraphics[height=3.5in,width=3.5in]%
{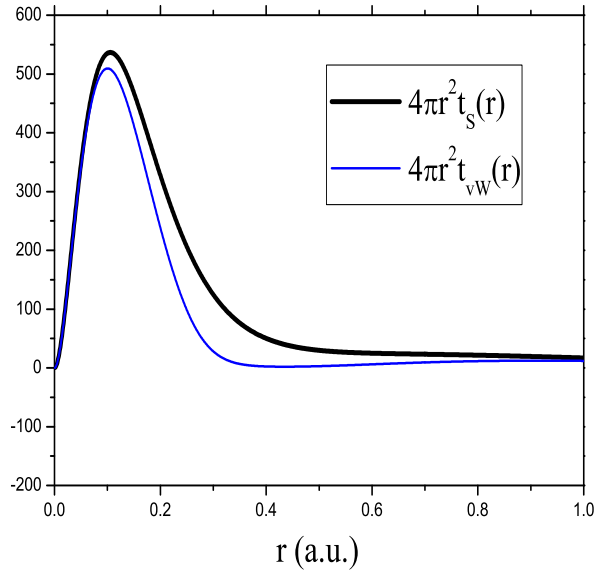}%
\caption{Radial vW kinetic energy densities for the neon atom.}%
\label{fig:NevW}
\end{center}
\end{figure}

In Figs.~\ref{fig:NeTF}, \ref{fig:NeGEA} and \ref{fig:NevW} 
we show the orbital-based KED for the Ne atom as a thick 
solid line, the approximated KEDs corresponding to the 
three functionals as dashed lines and the approximate KED
that includes the contribution due to the laplacian of the 
electron density is depicted with a thin solid line.

% -----------------------------------------------------

% -----------------------------------------------------

Being $t_{S}^{I}(\mathbf{r})$ always positive, 
we choose it for convenience as the reference KED,
whereas the corrections with the laplacians of the 
density are included as
\textquotedblleft laplacian contributions" 
to the approximated functionals.
This can be done because comparing 
$t_{S}^{L}(\mathbf{r})=%
t_{S}^{I} (\mathbf{r})+at_{0}(\mathbf{r})$ 
with $t_{S}^{func}(\mathbf{r})$ 
is equivalent to compare 
$t_{S}^{func}(\mathbf{r})-at_{0}(\mathbf{r})$ with 
$t_{S}^{I}(\mathbf{r})$. 
We use $t_{S}^{I}(\mathbf{r})$, always positive, 
because it shows more clearly the KED in different 
regions of the space. 
Specifically, the figures exhibit a clear 
first shell and a small shoulder at the second shell 
(these features are verified against the radial density, 
$4 \pi  r^2 n(r)$, and do not appear at the 
same positions when using  
$t_{S}^{II}(\mathbf{r})$).

For the TF functional (Fig.~\ref{fig:NeTF}) we see that, 
without the laplacian contribution, 
the KED distribution has its main contribution to 
the kinetic energy in a region nearer to the nuclei 
than $t_{S}^{I}(\mathbf{r})$ has. 
The approximate KED with the laplacian contribution 
yields a more similar distribution but 
two eye-catching pathologies are shown. 
The peak corresponding to the $1s$ shell is exaggerated and 
a defect of KED is found near the nucleus. 
Instead, the asymptotic behavior seems to be almost 
correct.

For GEA2 we can observe (Fig.~\ref{fig:NeGEA}) qualitatively 
almost the same behavior than the TF, 
but now the pathologies are bigger. 
It is now clear how this gradient correction yields better 
values of the total kinetic energy than TF, without 
improving the local behavior of the KED, 
being the differences with the reference distribution 
bigger and the value of $\sigma$\ worse.

For the von Weizs\"{a}cker functional (Fig.~\ref{fig:NevW}) 
very different results are obtained. 
The approximated KED is always smaller than the exact one, 
reflecting that when the von Weizs\"{a}cker functional is 
written in the form given by the equation (\ref{eq:vW}), 
it is a local lower bound for the first definition of 
the KED.
That is, the vW functional is not only a lower bound 
to the total kinetic energy but is also a local lower 
bound for the KED,\cite{1978HOHO}
and there is no need for any laplacian correction.
Note that in this case the contribution of the laplacian 
of the electron density yields a curve that is almost 
indistinguishable from the curve without it, 
as commented in section~\ref{sec:laplacian}, and it is 
not depicted. 
It can be noted that $t_{vW}$ is only correct near the 
nucleus and in the asymptotic decay -- the single orbital 
regions--, 
as expected from the definition of the vW functional.%
\cite{1996YLW}

%------------------------------------------------------------------------

\section{Conclusions}

We have developed a method to test kinetic energy density 
functionals attending not only to the total kinetic energies 
but looking at the local behavior of the kinetic energy 
densities associated to the functionals. 
To obtain quantitative measure of the accumulated difference 
between the distributions we have defined a quantity $\sigma$\ 
that we have called \emph{quality factor}. 
Due to the non-uniqueness of the KED definition we have
performed comparisons with an infinite family of kinetic 
energy densities, $t_{S}^{L}(\mathbf{r})$, 
generated by adding to the definition of the 
orbital-based KED the laplacian of the electron density 
multiplied by a variable prefactor.

The procedure have been employed to test the local quality 
of twenty one semilocal functionals. 
We check the fitting of the KED corresponding to each
functional to the closest $t_{S}^{L}(\mathbf{r})$ given by Eq.
(\ref{eq:tslapden}). 
For a given functional, we got values of $a$ (that weights 
the contribution of the laplacian), corresponding to a 
minimum value of $\sigma$, that show a small dependence 
on the atomic number~$Z$. 
And for that value of $a$ the corresponding KED is always 
closer to the mean of the first and second definitions 
of the orbital-based KED 
than to the definitions themselves. 
This result recalls that this mean has been proved to be 
the more natural definition for a
\textquotedblleft classical" KED.\cite{1991LMM,1996YLW}

The main result found is the unexpected failure of all 
semilocal functionals
but those with a full vW term to improve the local pathologies of the
Thomas-Fermi functional. 
Our measurement technique assures that, even the GGA
corrections to the TF functional always yield better 
total kinetic energies,
this semilocal functionals get worse local KEDs. 
This result confirms the
preliminar calculations presented in Ref.~\onlinecite{b2005GAA}. 
The only
exception to this rule is the Pearson functional, that yields 
bad values for the total kinetic energies.

Finally, we have qualitatively studied the behavior of the 
KEDs, showing the characteristic pathologies of the TF 
functional, that exhibits an excess in the first peak of 
the density (corresponding to the $1s$ orbital) and a defect
near the nucleus. 
This pathologies are always enlarged in  all the semilocal 
functionals, reflecting the main conclusion of this paper.

%------------------------------------------------------------------------

\begin{acknowledgments}
We acknowledge the continuous interest in this work of 
Prof. Rafael Almeida.
This work has been partially supported by a grant of 
the Spanish Ministerio de
Educaci\'{o}n y Ciencia (FIS2004-05035-C03-03).
\end{acknowledgments}

%------------------------------------------------------------------------
%Create the reference section using BibTeX:

\bibliography{Article_KEDSM_resubmission}

%------------------------------------------------------------------------

%\newpage

%------------------------------------------------------------------------
%TABLES ----------------------------------------------------------------
%------------------------------------------------------------------------

%
%\input{Tabla_Zeff.tex}
%
%
%\input{Tabla_ErrRelativo.tex}
%
%
%\input{Tabla_aLapDens.tex}
%
%
%\clearpage
%
%
%\input{Tabla_SigLapDens.tex}
%
%
%\input{Tabla_CAT.tex}
%
%\clearpage
%
%
%------------------------------------------------------------------------
%Figures (captions and .eps) -------------------------------------------
%------------------------------------------------------------------------

% Captions and Figures
%
%
%\begin{figure}[h!]
%\begin{center}
%\includegraphics[height=3.5in,width=4.25in]%
%{NeTF_KED.eps}%
%\caption{Radial TF kinetic energy densities for the neon atom.}%
%\label{fig:NeTF}
%\end{center}
%\end{figure}
%
 -----------------------------------------------------
%\begin{figure}[h!]
%\begin{center}
%\includegraphics[height=3.5in,width=4.25in]%
%{NeGEA2_KED.eps}%
%\caption{Radial GEA2 kinetic energy densities for the neon atom.}%
%\label{fig:NeGEA}
%\end{center}
%\end{figure}
%
%
 -----------------------------------------------------
%\begin{figure}[h!]
%\begin{center}
%\includegraphics[height=3.5in,width=4.25in]%
%{NevW_KED.eps}%
%\caption{Radial vW kinetic energy densities for the neon atom.}%
%\label{fig:NevW}
%\end{center}
%\end{figure}

%\input{Article_KEDSM_resubmission_Figures.txt}

%------------------------------------------------------------------------
%------------------------------------------------------------------------

\end{document}